\documentclass[twocolumn,preprintnumbers,amsmath,amssymb]{revtex4}
\makeatletter
\newcommand\footnoteref[1]{\protected@xdef\@thefnmark{\ref{#1}}\@footnotemark}
\makeatother
\usepackage{graphicx}
\usepackage{dcolumn}
\usepackage{bm}
\usepackage{ulem}

\begin{document}

\title{Probing the Casimir force with optical tweezers}

\author{ D. S. Ether Jr.$^{1,2,\dagger}$, L. B. Pires$^{1,2,\dagger}$, S. Umrath$^{3}$,  D. Martinez$^{1}$, 
Y. Ayala$^{1,2}$, 
B. Pontes$^{2}$, G. R. de S. Ara\'ujo$^{4}$,  
 S. Frases$^{4}$,  
 G.-L. Ingold$^{3}$, F. S. S. Rosa$^{1}$, N. B. Viana$^{1,2}$, H. M. Nussenzveig$^{1,2}$
 and P. A. Maia Neto$^{1,2,\footnote{pamn@if.ufrj.br}}$}

\address{$^1$Instituto de F\'isica, Universidade Federal do Rio de Janeiro, Caixa Postal 68528, Rio de Janeiro, RJ, 21941-972, Brasil}
\address{$^2$LPO-COPEA, Instituto de Ci\^encias Biom\'edicas, Universidade Federal do Rio de Janeiro, Rio de Janeiro, RJ, 21941-902, Brasil}
\address{$^3$Institut f\"ur Physik, Universit\"at Augsburg, Universit\"atsstra\ss e 1, D-86135 Augsburg, Germany}
\address{$^4$ Laborat\'orio de Ultraestrutura Celular Hertha Meyer, 
Instituto de Biof'\'{\i}sica Carlos Chagas Filho, Universidade Federal do Rio de Janeiro, Rio de Janeiro, RJ, Brazil}
\address{$^{\dagger}$ LBP and DSE contributed equally to this work and should be considered as first authors}

\begin{abstract}

We propose to use optical tweezers to probe the Casimir interaction between microspheres 
inside a liquid medium 
for geometric aspect ratios far beyond the validity of the widely employed proximity force approximation.
This setup has the potential for
 revealing unprecedented features associated to the non-trivial role of the spherical curvatures. 
For a proof of concept, 
we measure femtonewton double  layer forces between polystyrene microspheres at distances above $400$ nm  by employing very soft 
optical tweezers, with stiffness of the order of  fractions of a fN/nm. 
As a future application,
we propose to tune the 
 Casimir interaction between a metallic 
 and a polystyrene microsphere in saline solution 
 from  attraction to repulsion by varying the salt concentration. 
 With those materials, the screened Casimir interaction may have a larger magnitude than the 
 unscreened one. 
 This line of investigation has the potential for
  bringing together different fields
  including classical and quantum optics,  statistical physics and colloid science,
  while paving the way for novel quantitative applications of optical tweezers in cell and molecular biology. 

\end{abstract}
  

\maketitle

The last two decades have witnessed a remarkable progress in 
the understanding of the Casimir interaction \cite{Casimir48} between material surfaces. 
On the theoretical front,
the scattering approach 
 now allows one to derive exact results for different materials and geometries, either in \cite{Lambrecht06, Emig07, Rahi09,Milton10} or out of
  thermal equilibrium \cite{Bimonte09}, by capturing the electromagnetic reverberation of the vacuum or thermal fluctuations between 
  the interacting surfaces~\cite{JaekelReynaud91}. 
 When the surface geometry corresponds to 
  some simple symmetry,   the 
 scattering operators can be expanded  in a suitable basis and lead to explicit numerical results in a relatively simple way. 
 Examples  include parallel planes \cite{JaekelReynaud91}, two spheres \cite{Emig07,Rodriguez-Lopez11,Umrath15}, 
 parallel planar diffraction gratings \cite{Marachevsky08,Davids10,Intravaia12,Noto14,Guerout15}
 and a spherical surface 
 interacting either with a plane \cite{Emig08,MaiaNeto08,CanaguierDurand09,CanaguierDurandPRA10} or a  planar grating
  \cite{Messina15}. 
More numerically-oriented approaches, based on finite-difference methods, allow to derive results for even more general geometries \cite{McCauley11}, 
at the cost of computational time.
Such theoretical developments were strongly motivated by a number of precise Casimir force measurements, in the distance range from hundreds of nanometers to 
a few micrometers \cite{Lamoreaux97,Sushkov11,Chan01, Decca07, Decca07B,Mohideen98, Palasantzas08, Munday08,  Iannuzzi09, Munday09, Chevrier09,  Banishev13}. 

Since it is extremely hard to control the parallelism of two macroscopic planar surfaces 
at  sub-micrometer distances, most modern 
experiments employ the 
plane-sphere geometry. 
The sphere radius $R$ is always much larger than the distance of closest approach $L.$ Typically  the geometric aspect ratio is chosen to be
$R/L \sim 10^3$ or larger in order 
to have a large effective interaction area $\sim RL,$ thereby  raising the Casimir force to values $>10^2$ pN. 
Experimental techniques include the use of
 torsion pendulum apparatus \cite{Lamoreaux97,Sushkov11}, micro-machined oscillators \cite{Chan01, Decca07, Decca07B} 
and atomic force microscopes (AFM), 
 with the microsphere attached to the tip of the cantilever, operating either in
vacuum \cite{Mohideen98,  Palasantzas08,  Chevrier09, Banishev13}, in air \cite{Iannuzzi09} or in a liquid environment
 \cite{Munday08, Munday09}. 
 AFM has also been used to probe the total surface force, that includes a Casimir contribution, between spherical colloids in an aqueous medium \cite{Popa10, Wodka14}.

Although the scattering approach allows, in principle, the evaluation of the Casimir force in the plane-sphere geometry
by combining the spherical multipole and plane wave basis \cite{MaiaNeto08}, 
explicit results are still not available for the large aspect ratios
$R/L$
 found in the experimental conditions, 
since the multipole order required for numerical convergence 
scales as $R/L.$
Comparison between
theory and experiment 
 is thus entirely based on the
 Derjaguin or Proximity Force Approximation (PFA)~\cite{Israelachvili92}, in which the sphere curvature plays a trivial role.
Within PFA, the Casimir force is calculated from the result for parallel planes by a simple average over the local distances 
and diffraction is not taken into account. 
Although this approach is expected to provide accurate results for very large aspect ratios, as indicated experimentally in~\cite{Decca07B}, 
the magnitude of the leading-order correction to PFA for real experimental conditions is still unknown to this date, in 
spite of some recent advances~\cite{Fosco11, Teo11, Bimonte12}. 

Here we present an experimental proposal to measure the Casimir interaction between microspheres at moderate aspect ratios,  $R/L < 5,$
which is  within reach of the scattering approach based on the multipole expansion and 
well beyond the range of validity  of PFA. 
In this regime, the Casimir interaction results from the Mie scattering of vacuum and thermal field fluctuations between the 
spheres, with the spherical curvatures fully taken into account. For such geometries, the Casimir force is necessarily much weaker than the typical values 
probed in the PFA regime. Therefore, a very sensitive force probe  is required. 

Optical tweezers  \cite{Ashkin86,Ashkin06} allow for fN force measurements. 
Examples include the  force exerted by a single DNA molecule~\cite{Meiners00} and surface forces in colloids~\cite{Sainis07}.
One possibility is to use the optical trap only to control the initial position and measure the interaction potential between colloidal spheres from 
the statistical properties of the trajectories
with the trapping laser beam switched off \cite{Crocker94,Sainis07}. However, optical tweezers are more often employed as true
 force transducers. In this case, the force sensitivity benefits from the possibility of tuning the trap stiffness to very low values, which can be achieved 
 by reducing the 
trapping laser beam power. 
This is in sharp contrast with the AFM technique, where the cantilever stiffness is of course fixed for a given setup.

An important application of optical tweezers  is to measure 
the electrostatic double layer force between a microsphere and the 
plane coverslip surface at the bottom of the sample, along the direction parallel to the trapping beam \cite{Clapp01,Hansen05,Schaffer07}, as illustrated on the right side 
of fig.~1.
In this setup, the  trap stiffness varies with the sphere-surface distance, mainly because the spherical aberration introduced by refraction at the 
glass-sample interface depends strongly on the focal height \cite{Torok95, Viana07}. Moreover,  optical reverberation of the trapping beam is favored as the distance is reduced to the small values typically probed experimentally \cite{Schaffer07}. Thus, it is extremely hard to disentangle the desired signal from the trap perturbations as the 
microsphere approaches the coverslip surface.

Here we propose to measure the interaction force between two microspheres along a direction transverse to the trapping beam $z-$axis, as shown in
fig.~1.
In this configuration, interface
spherical aberration remains constant as the distance between the two microspheres is reduced. Moreover,   the trapping beam reverberation is negligible 
as long as the trapped sphere diameter is larger than the laser spot size, which is of the order of the laser wavelength for our diffraction limited spot. 
Thus, the force measurement can rely on a trap stiffness calibration for an isolated microsphere (see \cite{Ashkin06,Berg-Sorensen04} and references therein) since
the  optical trap potential 
 remains unchanged as we approach the second sphere along the transverse direction.     

Our two microspheres have different radii: microsphere $A,$ of radius $R_A,$ is optically trapped at a height equal to 
the radius 
$R_B>R_A$  of microsphere $B.$
The latter sits on the coverslip at the bottom of our sample, and its lateral position is controlled  by a nano-positioning stage, as discussed in more detail below. 
As we approach the larger sphere, the trapped sphere is displaced laterally from its equilibrium position and the force is obtained from the measurement of the 
lateral displacement once the trap stiffness is known. 
We calibrate the transverse trap stiffness by the Stokes-Fax\'en method \cite{Viana07} and find
 $k=0.26\pm 0.02$ fN/nm. Such an ultra-soft trap allows us to measure fN forces from  lateral displacements of a few nm, within  reach of our spatial resolution. 
Double-layer forces  in the pN range have been measured in a similar setup, 
with a stiffness of 85 fN/nm,
by holding one of the beads with the help of a micropipette~\cite{Gutshe07}.

In order to demonstrate
 the feasibility of our proposal, we measure the double-layer force between two polystyrene beads in ultrapure water, at long distances well beyond the validity of PFA. As discussed in detail below, our sample is sufficiently clean so as to lead to very large Debye screening lengths, thus  resulting in  fN forces at distances above $400\,{\rm nm}.$  The experimental setup is sketched in fig.~\ref{fig:setup}. A linearly polarized collimated laser beam 
(IPG photonics, model YLR-5-1064LP) with wavelength $\lambda_{0}=1064\, {\rm nm}$ and waist $w_{0}=(2.84 \pm 0.02)\,{\rm mm}$ propagates through a half-wave plate (HWP) and a polarizing beam splitter (PBS),  
 enabling us to control  the transmitted beam power $P\sim 80\,{\rm mW}.$ 
 After crossing 
 an absorptive neutral density filter (ND, Thorlabs NE13A D-1.3), the attenuated beam ($P\sim 4\,{\rm mW}$) is divided by a 50:50 beam splitter. The reflected beam
 allows for real-time  power monitoring with the help of an optical power meter. The transmitted beam is reflected by a dichroic mirror and focused into our sample after crossing the oil-immersion objective lens (Nikon PLAN Fluor, 100x, NA 1.3, $\infty$/0.17, WD=0.20, type A $n=1.515$ immersion oil) of a Nikon TI-U  inverted microscope.
 The beam power  at the objective entrance is $P\sim1\, {\rm  mW}.$
Our  sample is
 attached to a piezoelectric nano-positioning stage (Digital Piezo Controller E-710, Physik Instrumente). Finally, the
  incoherent condenser light   scattered from the
  sample is collected by the objective lens, transmitted through the dichroic mirror,   collected again by a lens L and captured by  a CMOS camera (Hamamatsu Orca\textregistered-Flash2.8 C11440-10C) for data analysis.

 Our sample consists of a colloidal dispersion contained between 
 glass coverslip  surfaces (Knittel, Germany) 
 as shown in the inset of fig.~\ref{fig:setup}.
 The surfaces are
 cleansed by 
 immersion in a sulfonitric mixed acid solution (nitrating acid mixture) for one hour, then rinsed 30 times with Milli-Q water, and finally stored in absolute ethanol.
  Prior to the experiments, the coverslips are dried inside a biosafety cabinet to avoid contamination, and then assembled one on top of the other using an industrial adhesive (ELASTOSIL\copyright E43 Wacker) to build the confining cell depicted in the inset of fig.~\ref{fig:setup}. 
  The  uncoated polystyrene beads (Polysciences, Inc.) used in the experiment have 
   radii
   $R_{A}=\left(1.503\pm 0.006\right) {\rm \mu m}$ and $R_{B}=\left(7.18\pm0.09\right) {\rm \mu m},$
   as measured by electron microscopy,
  where the uncertainties are the standard errors.
  These microspheres are diluted in Milli-Q water, and the prepared solution is injected into the sample, which is then sealed with the industrial adhesive.

\begin{figure}\centering  
\includegraphics[scale=0.315]{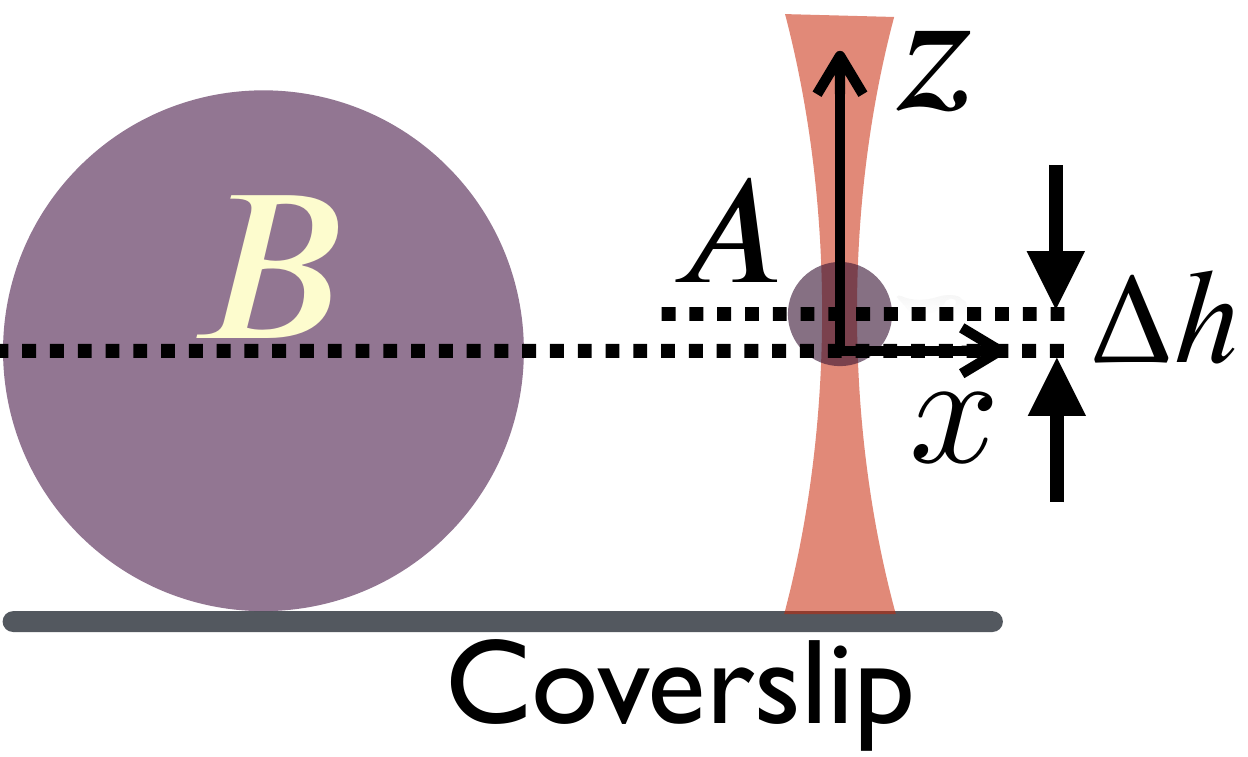}
\caption{\label{fig:scheme} (Color online) 
Scheme for measuring the Casimir force: 
as microsphere $B$ approaches the laser beam axis, the optically trapped microsphere $A$ is displaced laterally. The initial misalignment $\Delta h$ 
is measured by defocusing microscopy and then corrected by displacing the microscope stage.
}
\end{figure}

A crucial experimental issue is the alignment between the two sphere centers. 
The alignment on the $xy-$plane is easily achieved by using the information from the recorded images as an input for driving the piezo nano-positioning stage.
The  alignment along the $z-$axis, however, is more challenging, since it requires to control the axial position of the trapped bead $A.$ 
 We combine 
defocusing microscopy~\cite{Agero03}
and the axial positioning procedure described in~\cite{Nicholas14}. 
Our method is based on measuring the contrast $C_{A(B)}=I_{A(B)}/I_{\rm back}-1,$ where $I_{A(B)}$ and $I_{\rm back}$ are the
intensities at the center of sphere $A$ ($B$) image and at the  background, respectively. 
The large bead $B$ can be modeled as a phase object, for which $C_B=0$ exactly when the object  plane (conjugate of the image plane at the CMOS camera) coincides with its equator plane,  thus defining the reference for the measurement of the trapped bead axial position $\Delta h,$ as shown in fig.~\ref{fig:scheme}.

Since  bead $A$ 
is too small to be modeled as a phase object, we need to calibrate its contrast as function 
of its position with respect to the object plane. 
In order to control the position, we turn off the trap, attach bead $A$ 
 to the coverslip  (not shown), and record its contrast as the stage is displaced vertically. We can identify the stage position for which the object and coverslip planes coincide. Once we know this reference, 
we can use the calibrated  contrast function $C_A$ to determine the axial equilibrium position $\Delta h$  from the measured contrast value when bead $A$ is optically trapped. 

Alternatively, we 
can also measure the contrast of the trapped bead as we displace the stage upward.  In this case, the contrast is initially constant, because the trap equilibrium position is fixed with respect to the object plane. However, once the coverslip starts to push the bead out of the trap, the measured contrast starts to follow the calibrated function
$C_A,$ thus defining again a reference plane. The values for $\Delta h$ resulting from the two alternative procedures coincide: $\Delta h=\left(0.4\pm0.2\right)\mu {\rm m}.$
Finally, we compensate for the initial misalignment  by displacing the stage vertically.
We test our result by measuring the center-to-center distance when the sphere surfaces touch: from video microscopy we find 
$R_A+R_B  =  (8.72 \, \pm\,  0.04)\, \mu{\rm m},$
 in agreement with the result $(8.68 \pm 0.09)\, \mu{\rm m}$
found by adding the values measured  with electron microscopy.

\begin{figure}\centering  
\includegraphics[scale=0.295]{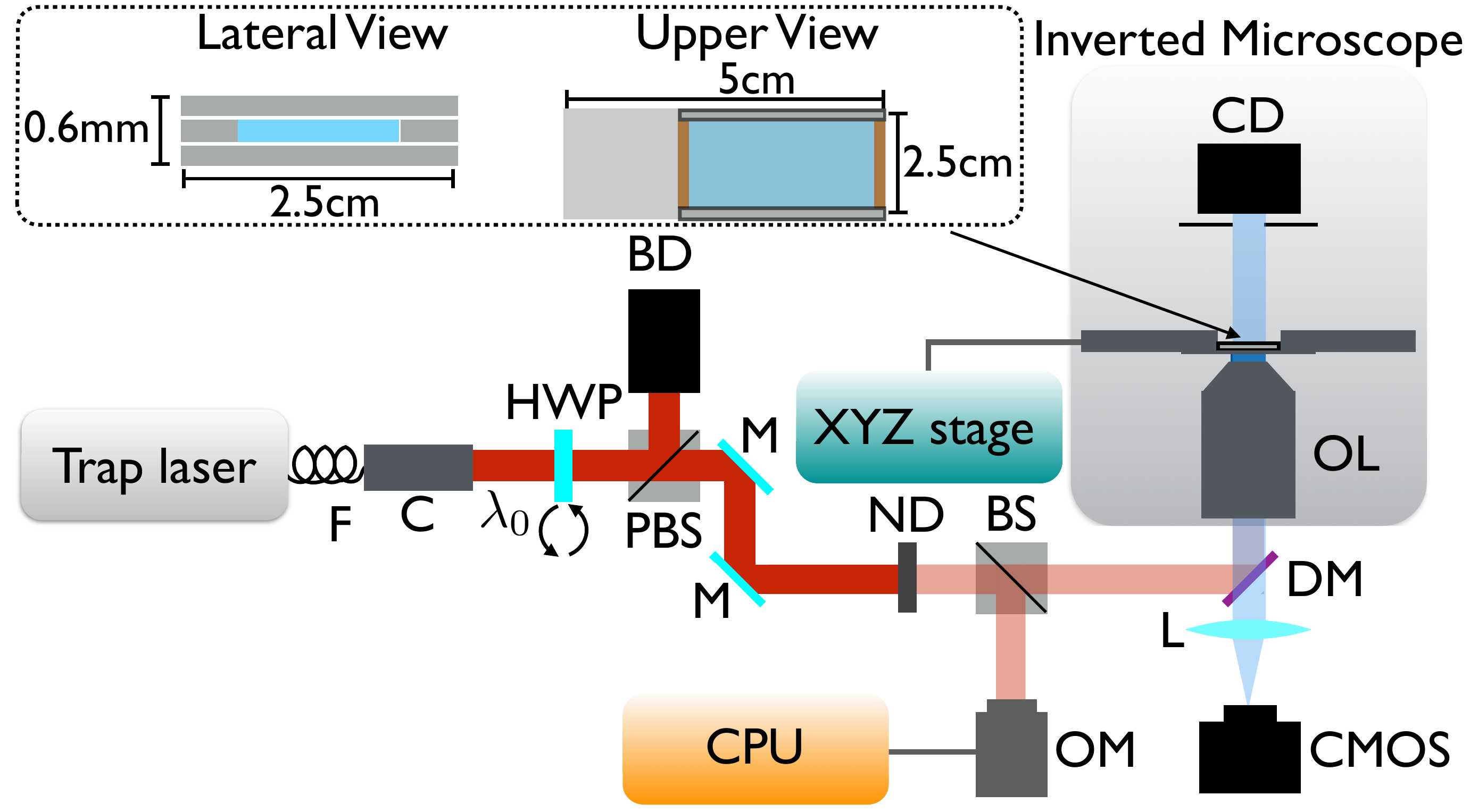}
\caption{\label{fig:setup} (Color online) Experimental sketch. BS: beam splitter; BD: beam dump; C: beam collimator; CD: condenser; CMOS: camera; DM: dichroic mirror; F: single-mode polarization-maintaining optical fiber; HWP: half wave plate; L: lens; M: mirror; ND: neutral density filter; OL: objective lens; OM: optical power meter; PBS: polarizing beam splitter; XYZ stage: piezoelectric translational stage. Inset: sample cell.}
\end{figure}

Once the microspheres are aligned, we set the microscope stage into motion along the $x-$direction
with velocity $v=15\,\mbox{nm/s}$ and record 
images of sphere $A$ with a repetition rate of  $667\, \mbox{frames/s}$  during $\sim 3\,{\rm min}.$ From those images, 
we compute the  lateral displacement $x(t)$ of the sphere from its unperturbed equilibrium position, as defined by the beam axis,   as function of time $t$. 
 We repeat this procedure
 20 times. 
 
 There are three different regimes for each experimental run: (i) When sphere $B$ is sufficiently far, sphere $A$ undergoes Brownian fluctuations around the equilibrium position close to the beam axis. The 
  downside of  having a very low trap stiffness $k$
  is the enhancement  of Brownian fluctuations, whose lower bound is given by the 
 thermal equilibrium
 value at room temperature
 $x_{\rm rms}=\sqrt{k_{\rm B} T/k} \sim 0.13 \,\mu{\rm m}.$ Thus, we expect 
  the equilibrium regime to be limited to  
 distances  larger than $x_{\rm rms}.$ (ii) At some critical distance $\stackrel{>}{\scriptscriptstyle\sim}x_{\rm rms},$ 
 the combined stochastic and attractive forces overcome the trapping force and
 bead $A$  
 escapes from the trap, and (iii) 
 finally  gets attached to sphere $B.$ 
 
Fig.~3a
shows  the position  $x(t)$  for a single run, with the sphere $B$ approaching the beam axis as time $t$ increases. 
 We  zoom in for a detailed analysis 
of the region where the two spheres get close  to each other, corresponding to the final part of the run.  
In the figure, 
we can identify the three separate regimes discussed above: initially, the trapped bead $A$ 
undergoes Brownian fluctuations around its displaced equilibrium position. 
For intermediate times, indicated by the dotted ellipse in fig.~3a and corresponding to distances close to $400\,{\rm nm},$
bead $A$ is escaping from the trap and moving  towards bead $B.$ 
Finally, 
the third region shown in the rightmost part of the figure corresponds to the situation where bead $A$ is already attached to bead $B$ and simply follows the  microscope stage motion.

In order to understand the escape from the trap in a more quantitative way, we plot in the upper 
inset of fig.~3a the total potential energy of the trapped bead $A\! :$ 
  $U = U_{\rm opt}+U_{\rm DL}+U_{\rm Cas},$ 
where $U_{\rm opt},$ $U_{\rm DL}$ and $U_{\rm Cas}$ are  the optical trap, double layer and Casimir potentials, respectively (see below for details).
As the distance $L\rightarrow 0,$
 the attractive Casimir interaction
  dominates, so that  $U$ develops a potential barrier whose height is very sensitive to the distance $d$ between sphere $B$ and 
 the beam axis (see lower inset in fig.~3a).  
 The solid line corresponds to the shortest distance, 
$d=0.35\, \mu{\rm m}+R_A;$ the dotted line to $d=0.4\, \mu{\rm m}+R_A;$ and the dashed line to  $d=0.5\, \mu{\rm m}+R_A.$
  The barrier height decreases very fast as bead $B$ approaches the beam axis and
 becomes of the order of
  the thermal energy $k_{\rm B} T$
for $d-R_A<0.4\,\mu{\rm m},$ leading to a thermal diffusion above the barrier in this distance range. This is in agreement with our experimental observation
that bead $A$ escapes from the trap by moving  a distance $\sim 0.4\,\mu{\rm m}$ towards bead $B,$ as shown in the region highlighted by an ellipse in fig.~3a. 

The fact that the bead escapes from the trap limits our force measurement by the equilibrium position method to distances larger than $0.4\,\mu{\rm m}.$
Within this method, our signal is the displacement of the equilibrium position, which is also visible in the potentials shown in the inset of 
fig.~3a. The unperturbed equilibrium position is at $L=d-R_A,$ corresponding to the sphere $A$ centered at a position on the beam axis, and indicated by a vertical line 
for each of the three values of $d$ considered in fig.~3a.   
With respect to this position, the new equilibrium position is displaced to the right by an increasing amount  as
$d$ decreases,
due to the repulsive nature of the double layer interaction.

\begin{figure}\centering    
\includegraphics[scale=0.5]{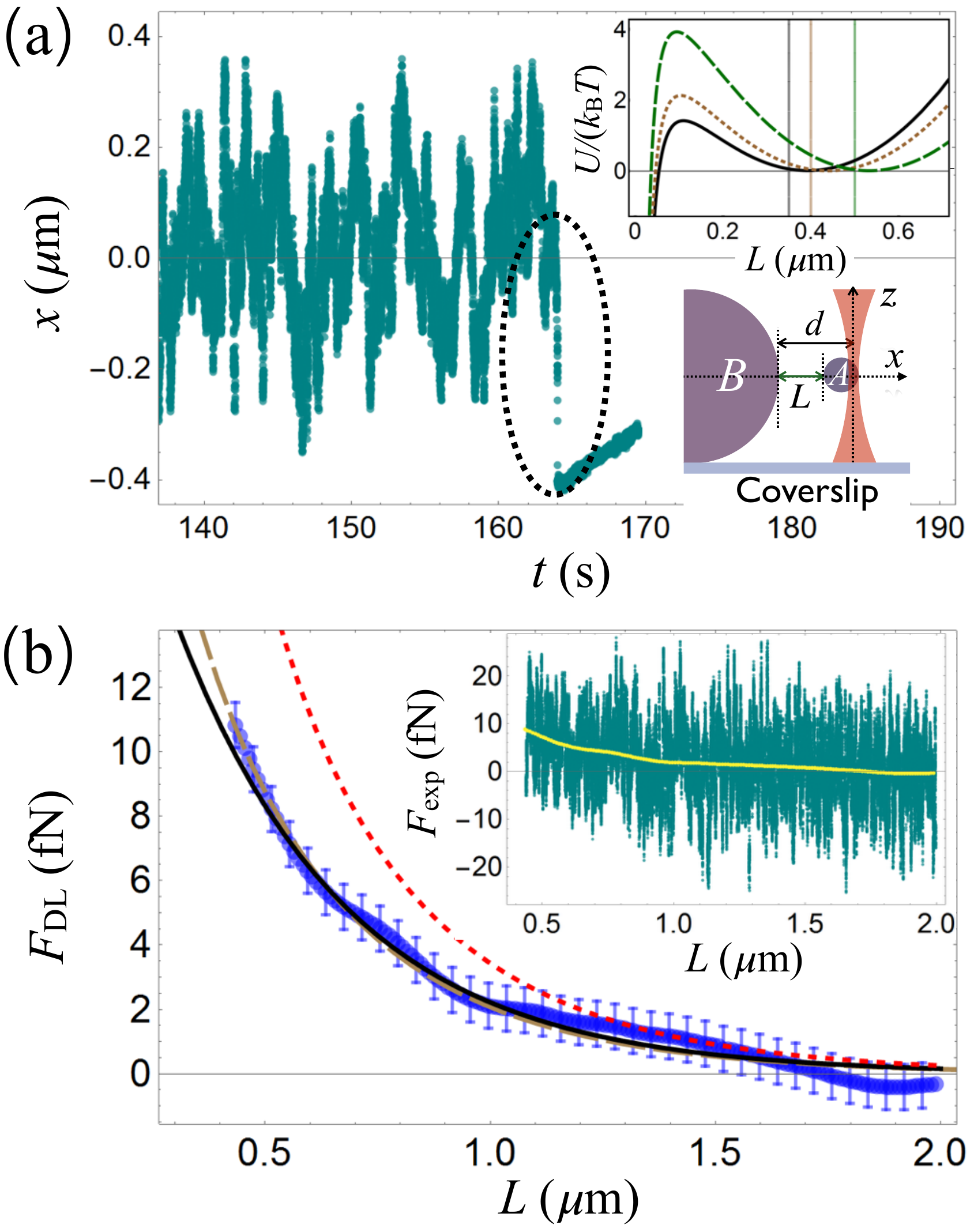}
\caption{\label{results} (Color online) Experimental data for the 
interaction force between two polystyrene beads in water. (a) Transverse position $x$ versus time 
corresponding to the final part of a single run, with $x=0$ representing the unperturbed equilibrium position at the intersection with the beam axis. 
The bead escapes from the trap in the 
region indicated by a dotted ellipse.
 Upper inset:  potential variation  (in units of $k_{\rm B} T$) with the surface-surface distance $L,$  for different transverse distances $d$ between the beam axis and the  sphere $B$ surface. 
 Solid: $d-R_A=0.35\, \mu{\rm m};$ dotted: $d-R_A=0.4\, \mu{\rm m};$ and dashed: $d-R_A=0.5\, \mu{\rm m}.$
 The vertical lines indicate the unperturbed equilibrium positions at $L=0.35\, \mu{\rm m},0.4\, \mu{\rm m}$ and $0.5\, \mu{\rm m}$ from left to right.
   Lower inset: definitions of  $L$ and $d$ (not drawn to scale).
(b) Double-layer force variation with $L.$
Dots: experimental data; solid: LSA fit (see text); long dash: exact solution of the linear Poisson-Boltzmann equation; short dash: PFA theory  for the double layer force. 
 For clarity not all error bars are represented. Inset: raw  data for the total force, obtained from the average over 20 runs.
 Full line: result after smoothing (see text).
   }
\end{figure}

 As discussed below, the Casimir force between our polystyrene spheres in water is too weak to be measurable in the distance range 
 probed experimentally. 
 Our signal is then dominated by the  double-layer force $F_{\rm DL},$ 
 which is obtained by subtracting the Casimir force from 
the experimental data: $F_{\rm DL}= F_{\rm exp}- F_{\rm Cas},$ with $F_{\rm exp}= k x.$
The net force $F_{\rm exp}$ is repulsive, so that the average bead displacement $x$ is positive. 
We smooth our raw data set, representing an average over 20 runs and shown as an inset of fig.~\ref{results}b,
by the locally weighted least square error method, which
reduces the contribution from outliers associated to rare large fluctuations.
We estimate our experimental uncertainty
from the standard error of the mean $\delta x$ and the stiffness uncertainty   $\delta k$:
$\delta F_{\rm DL}=\delta F_{\rm exp}= F_{\rm exp}\sqrt{(\delta k/k)^2+(\delta x/x)^2}=0.7\,{\rm fN}$
when taking the shortest distance in our data set. 
  In fig.~3b, we plot the resulting  double-layer force  versus distance  together with different theoretical models.
 All of them correspond to the linear Debye-H\"uckel approximation of the Poisson-Boltzmann equation \cite{Butt}, valid when the 
 electrostatic energy per ion is much smaller than $k_{\rm B} T.$ We also assume  the  charge densities on the polystyrene surfaces to be uniform, 
 independent of distance and the same for both microspheres. Both approximations are valid for the large distances probed in our experiment \cite{Carnie94} and are consistent with our 
 experimental results~\cite{footnote_regulation}.

 We use the 
  linear superposition approximation (LSA) \cite{Bell70}
  to fit the experimental data in the 
  range $0.6\, \mu{\rm m}\le L \le 1.0 \, \mu{\rm m}$
  and find $\lambda_D= (418 \pm 7)\,{\rm nm}$ for the 
  Debye screening length, compatible with the interval expected for a sample of ultrapure water~ \cite{Butt},  
   and $\sigma=(1.88 \pm0.04)\, \mu{\rm C/m^2}$ for the surface  charge density. The solid line 
   in  fig.~\ref{results}b  represents the resulting LSA force variation. 
LSA provides an accurate description
  for distances larger than
  the screening length, so that the double layer around each sphere is approximately unaffected by the presence of the other one. 
  Deviations from LSA are expected as the distance decreases, which is suggested by fig.~\ref{results}b.
  The long-dashed line represents the exact solution of the 
  linear Poisson-Boltzmann equation~\cite{Carnie94} taking the boundary conditions corresponding to our two microspheres into account \cite{foot_exact}, and using the parameters found from the LSA fit. Comparison with our experimental data confirms that our exact solution captures 
  the effect of deviation from LSA, although our experimental precision is not sufficient to resolve the difference in a more definite way.
  
  Fig.~\ref{results}b  also shows that the PFA (short dash), again calculated with the parameters from the LSA fit, grossly
  overestimates the double-layer force for the parameters corresponding to our experiment. 
  This is expected, since the PFA is particularly poor when describing long range interactions and thus requires  Debye screening lengths much shorter 
  than both radii, a condition not met in our experiment. 
  
Our sensitivity  at the fN  level  opens the way for future Casimir force measurements
 at long distances, which would also be well beyond the scope of PFA.
We have derived theoretical results for different materials, in order to understand which Casimir experiments would be feasible with 
our current setup. For that purpose, we use the scattering approach for the geometry corresponding to two spheres \cite{Emig07,Rodriguez-Lopez11,Umrath15}.
At finite temperature, the Casimir force is given by a sum over the  Matsubara frequencies $\xi_n\equiv \mbox 2\pi n k_{\rm B}T/\hbar, \; n\ge 0$ \cite{Lambrecht06}.
We discuss two limiting cases: for large Debye  lengths, $L\ll \lambda_D,$
 we can neglect all screening effects and then perform  the  Matsubara summation as if there were no ions in solution. 
In the opposite limit, $ \lambda_D\ll L,$ the zero-frequency contribution is completely screened by the ionic charge distribution in the intervening liquid
and can, therefore, be neglected in the derivation of the Casimir force~\cite{Parsegian}.

We first discuss the case of the experiment reported above, with two polystyrene microspheres in water. 
 The dielectric functions of 
  polystyrene $\epsilon_{\rm Ps}$  and water
  $\epsilon_{w},$ to be 
   evaluated at the imaginary frequencies $\mbox{i}\xi_n,$ 
   are described by Lorentz models. 
   We take the
 parameters given by~\cite{vanZwol10}
 and, for water, modify   the dielectric function in order to
 account for the correct zero frequency value $\epsilon_w(0)=78.7.$ 

 In fig.~4, we plot the magnitude of the attractive Casimir force versus distance, at room temperature $T=293\, \mbox{K}$
 with no screening (top) and with complete suppression of the zero-frequency contribution (bottom), which corresponds  to the limit of strong screening. The former is used to compute the total potential $U$ shown in the upper inset of fig.~3a and also to extract the double-layer 
 force from the experimental total force shown in fig.~3b.
 The third curve in-between represents the 
 zero-temperature case. The plot shows that the force is indeed too weak to be measurable in the range $L>400\,{\rm nm}.$
   Nevertheless, it  displays some remarkable properties, that could possibly be tested experimentally by taking larger spheres. 
 The unscreened Casimir force is dominated by the zero Matsubara frequency contribution
 because of the large value of $\epsilon_w(0).$ 
  There are two consequences:   First,   the
  thermal contribution overwhelms the zero-temperature  vacuum contribution even at  relatively short distances, of the order of $0.1\,\mu{\rm m},$
  as clearly  shown in fig.~4. 
  In fact, the force can be approximated by the  
 high-temperature classical result, proportional to $T$ \cite{Canaguier-Durand12},  
 at such distances. 
 Second,  the magnitude of the force can in principle be tuned
to any value between the top and bottom curves in fig.~4, corresponding to a variation of at least one order of magnitude,
 by selecting a suitable value for the Debye screening length $\lambda_D.$
 This can be achieved by changing the salt concentration of the colloidal suspension.  
However,  since 
the Casimir and the double-layer forces 
 are screened in essentially the same way in this case, 
  it would be  hard to experimentally isolate the former from the total surface force. This is in contrast to
   the configuration with mercury micro-droplets to be discussed
   in the final part of this letter, 
 in which the suppression  of the zero frequency contribution 
  might actually lead to a larger force magnitude.

\begin{figure}\centering  
\includegraphics[scale=0.41]{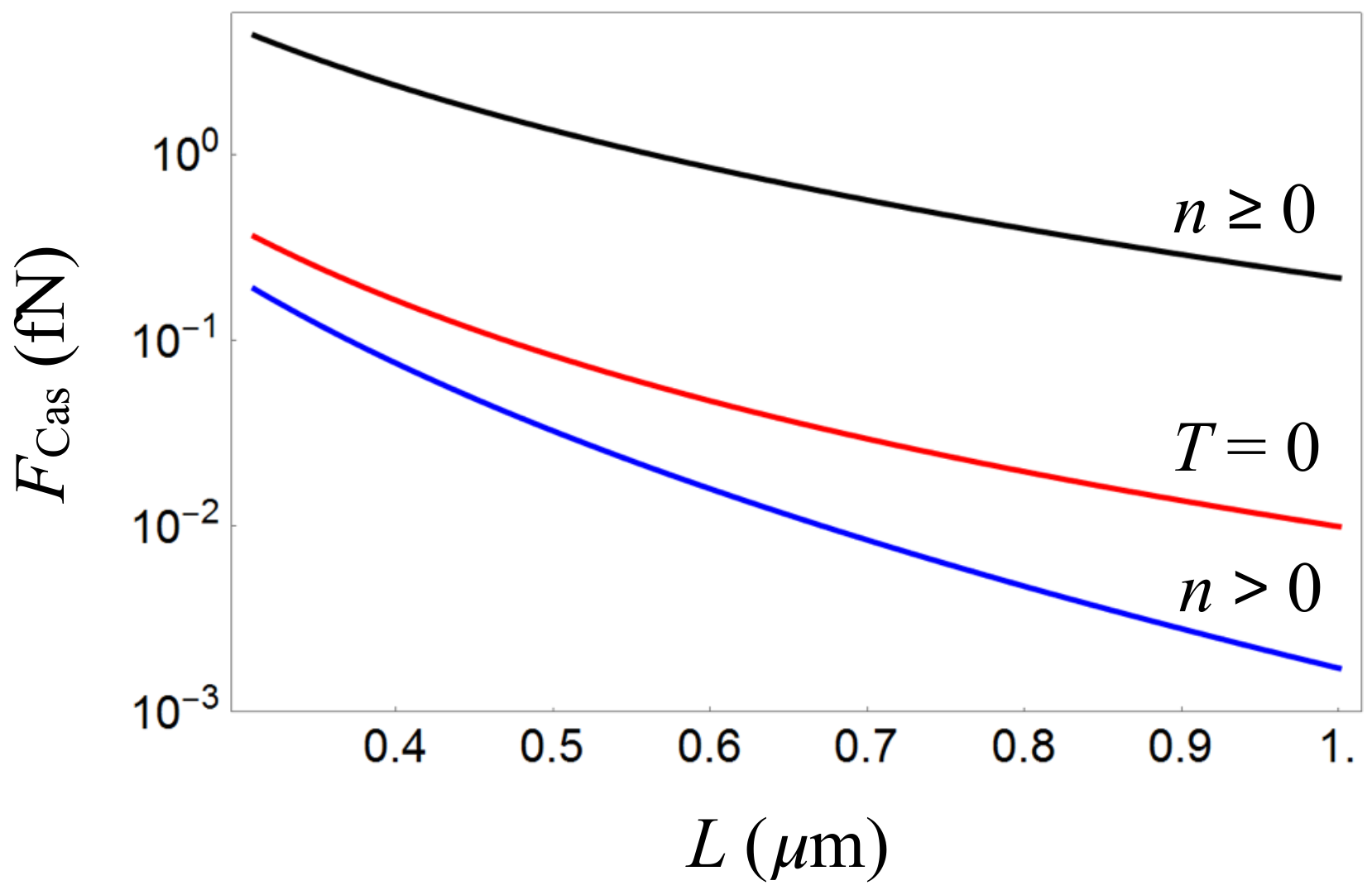}
\caption{\label{figt1} (Color online) 
Theoretical Casimir force magnitude between two polystyrene microspheres in water 
versus distance. We take the experimental values for the radii: $R_A=1.503\,\mu{\rm m}$ and $R_B= 7.18\,\mu{\rm m}.$
From top to bottom: finite temperature $T=293\, \mbox{K}$ with all Matsubara frequencies $\xi_n \; n\ge 0$ taken into account, representing the case of no screening;
$T=0;$ and $T=293\, \mbox{K}$ excluding the zero-frequency contribution $n=0,$ representing the situation with strong screening.
}
\end{figure}

\begin{figure}\centering  
\includegraphics[scale=0.42]{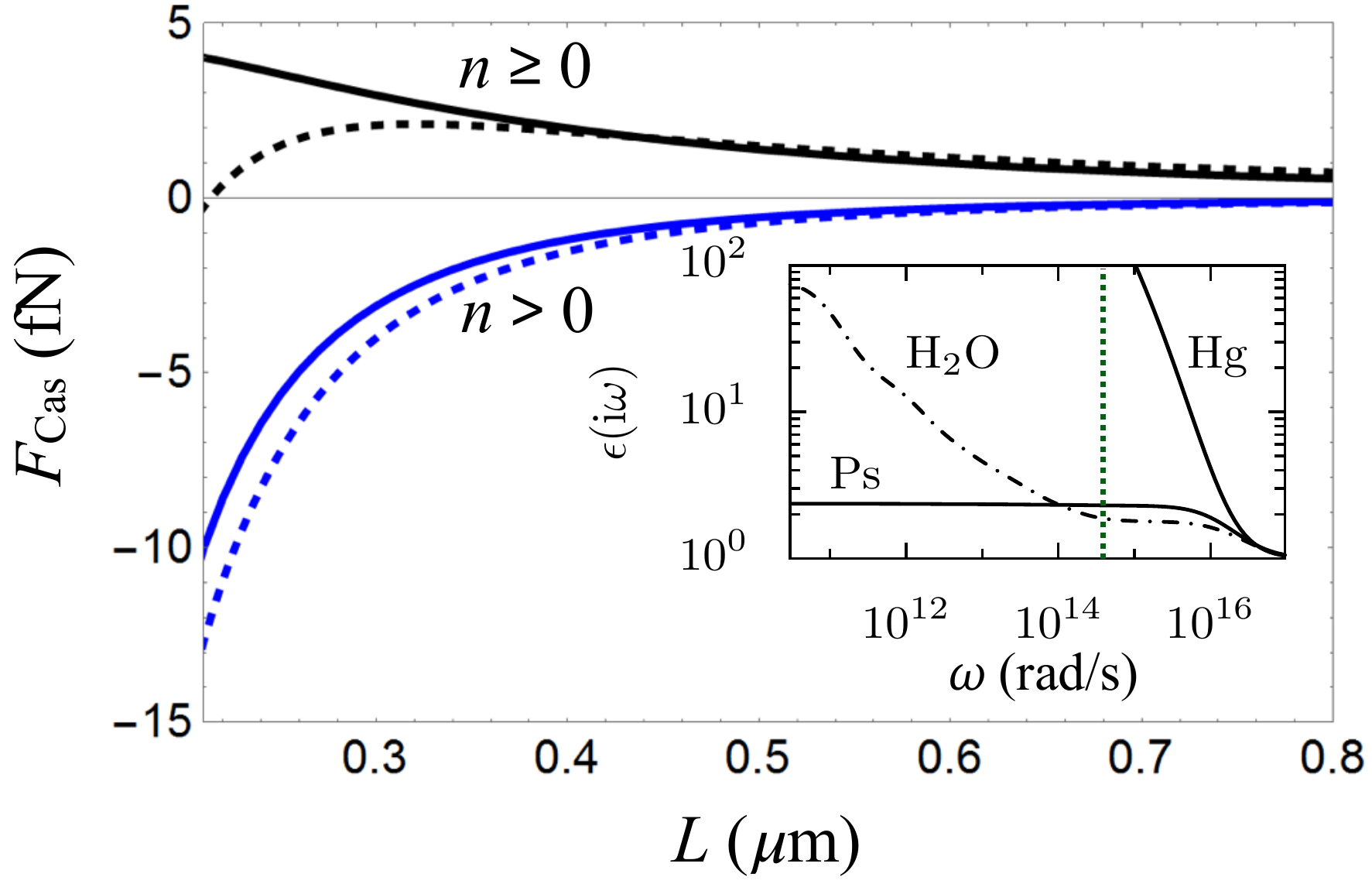}
\caption{\label{figt2} (Color online) 
Theoretical Casimir force between 
a polystyrene microsphere of radius $R_A=2\,\mu{\rm m}$ and 
a mercury droplet of radius  $R_B= 7\,\mu{\rm m}$ in water
versus distance. 
Same conventions as in fig.~4, with a positive (negative) sign representing repulsion (attraction). We also show the PFA results (dashed lines) for both unscreened and screened cases. 
Inset: dielectric functions of mercury, water and polystyrene on the imaginary frequency axis. For convenience, we indicate the first positive Matsubara frequency at room temperature
by a vertical dotted line. 
}
\end{figure}

 Mercury micro-droplets  can be produced with our setup
 in a relatively simple way \cite{Viana02}. 
 Because of the surface tension effect,  they are very spherical, with no surface roughness. In addition, patch effects should be 
reduced with respect to the  standard case of solid metallic surfaces \cite{Esquivel-Sirvent14}. 
We compute the Casimir force between a mercury 
droplet of radius $R_B=7\,\mu{\rm m}$ and a polystyrene bead of radius $R_A=2\,\mu{\rm m}$ inside an aqueous medium. 
 We take the Smith modification~\cite{Smith01} of the Drude model as detailed in~\cite{Esquivel-Sirvent14} to model the dielectric function of mercury
  $\epsilon_{\rm Hg}.$

The resulting Casimir force values are shown in fig.~5, again in the two limiting cases of no screening (upper curves) and strong screening (lower curves). 
In addition to the exact results, represented by the solid lines, we also plot the PFA values as dashed lines. 
In order to understand these results, we plot 
the dielectric functions of the three materials as an inset. At zero frequency, the permittivities are ordered so as to lead to a repulsive contribution \cite{DLP}, since 
 $\epsilon_{\rm Hg}> \epsilon_{w}>\epsilon_{\rm Ps},$  whereas all positive Matsubara frequencies yield attractive contributions, as illustrated by the first positive Matsubara frequency
 $\xi_1$
 indicated in the inset of fig.~5. The net Casimir force then results from 
 a delicate balance between these two opposite contributions (see also \cite{Rodriguez10} for related configurations). 
 Within PFA, they cancel each other at a distance slightly larger than $0.2 \,\mu{\rm m},$ where there is a crossover from repulsion to attraction, as shown by the upper dashed
  line in fig.~5. 
 On the other hand, when considering the full Mie scattering from each sphere, the zero frequency turns out to be dominant, and 
the net Casimir force is repulsive 
 for the whole range of distances shown in the figure (upper solid line), as long as 
 screening is negligible. Thus, even the qualitative features of the Casimir interaction in this configuration
  are not correctly described by the PFA. 
 
 The delicate balance between the repulsive and attractive effects
 is also modified by
 screening of the 
 zero frequency contribution.
 The lower curves  in fig.~5 correspond to the case of strong screening, with total suppression of 
 the zero frequency contribution, thus leading to an attractive force. 
 As a consequence, one could tune the interaction from repulsion to attraction by decreasing the Debye screening length. 
 
 The parameters and orders of magnitude corresponding to fig.~5 
 indicate that a Casimir force measurement should be feasible 
  with the polystyrene sphere $B$ replaced by a mercury droplet of similar size. 
 In an experiment currently under way, 
 we
tune the Debye screening length to $\lambda_D\sim 0.05\, \mu{\rm m}$ and measure the total interaction force in the range $>0.2 \,\mu{\rm m}.$
In this case, the Casimir force can be approximated by the lower solid curve in fig.~5 representing the limit of strong screening, 
with a magnitude similar to the double-layer forces we have measured.
In the regime of strong screening,
$\lambda_D\ll L,$ the Casimir interaction is singled out because 
 the double-layer force is completely suppressed \cite{Munday08}. 
  Therefore, it is  fortunate that in this limit
   the Casimir force has a larger magnitude than in the un-screened one  for distances $L\stackrel{<}{\scriptscriptstyle\sim} 0.3  \,\mu{\rm m},$
as illustrated by fig.~5.

In conclusion, optical tweezers are promising tools to probe the Casimir interaction for geometric aspect ratios far beyond the validity of the proximity force or  
Derjaguin approximation. 
 In our proposal,
the trap stiffness remains unchanged as we approach the second sphere along a direction perpendicular to the beam axis. 
In fact, since the focal height is kept fixed, so is the effect of spherical aberration on the optical force field. Moreover, optical
reverberation of  the trapping laser beam is negligible since 
its sub-micrometer spot size  is smaller than the sphere diameter.  
An important ingredient in our scheme is the axial alignment procedure, which is based 
on calibrating the trapped sphere diffraction pattern with the help of defocusing microscopy.

We have performed 
fN force measurements in a  simpler scenario
than that corresponding to  cell biology experiments, 
allowing us to check experimental procedures by comparison with relatively simple and accurate theoretical models for the double-layer force.
Such comparison still depends on  fitting parameters governing the interaction. 
On the other hand, the Casimir interaction is of a more universal nature, so that by applying our methods to Casimir force measurements it would be possible to perform blind theory-experiment comparisons. This opens the prospect for
 the development and tests of 
new force measurement strategies at the fN scale based on the use of optical tweezers, 
 with potential
 applications 
 in cell and molecular biology.

\acknowledgments

The authors are grateful to 
Ricardo Decca, Carlos Farina, Romain Gu\'erout, Astrid Lambrecht, Oscar Mesquita, Jeremy Munday, George Palasantzas and Serge Reynaud
 for  discussions. This work was partially funded  by 
  DAAD (Germany) and CAPES (Brazil) through the PROBRAL program,  and by
CNPq  and FAPERJ (Brazil).


\begin{thebibliography}{0}
\bibitem{footnote_regulation} 
For distances shorter than the Debye screening length, the charge density of each microsphere is modified by the presence of the other microsphere.
This effect is described by the regulation model \cite{Popa10,Carnie94}, whose results at larger distances coincide with those obtained by the simpler constant charge model considered here.
\bibitem{foot_exact} For that purpose, we have generalized the formalism presented in~\cite{Carnie94} in order to take the finite permittivity of polystyrene into account.
\end{thebibliography}
\end{document}